%% file: 0_AIC_Main.tex
\documentclass[
]{ceurart}

\usepackage{listings}
\usepackage{latexsym}
\usepackage{amssymb}
\usepackage{amsmath}
\usepackage{amsthm}
\usepackage{booktabs}
\usepackage{enumitem}
\usepackage{graphicx}
\usepackage{color}
\usepackage{subcaption}  % In your preamble

\begin{document}

\copyrightyear{2025}
\copyrightclause{Copyright for this paper by its authors.
  Use permitted under Creative Commons License Attribution 4.0
  International (CC BY 4.0).}

\conference{None}

\title{Robust Symbolic Reasoning for Visual Narratives via Hierarchical and Semantically Normalized Knowledge Graphs}

% \tnotetext[1]{You can use this document as the template for preparing your
%   publication. We recommend using the latest version of the ceurart style.}

% \author[1]{Yi-Chun Chen}[%
% orcid=0009-0003-4035-9894,
% email=yi-chun.chen@yale.edu \and ychen74@alumni.ncsu.edu
% url=https://sites.google.com/view/rimichen-web/bio
% ]
% \cormark[1]
% \address[1]{Yale University,
%  New Haven, CT, 06510, USA}

\author[1]{Yi-Chun Chen}[
    orcid=0009-0003-4035-9894,
    email={yi-chun.chen@yale.edu, ychen74@alumni.ncsu.edu},
    url=https://sites.google.com/view/rimichen-web/bio
]
\cormark[1]
\fnmark[1]

\address[1]{Yale University, New Haven, CT, 06510, USA}

% \author[1]{anonymous authors}

% \address[2]{Joint Institute for Nuclear Research,
%   6 Joliot-Curie, Dubna, Moscow region, 141980, Russian Federation}

% \author[3]{Ilaria Tiddi}[%
% orcid=0000-0001-7116-9338,
% email=i.tiddi@vu.nl,
% url=https://kmitd.github.io/ilaria/,
% \address[3]{Vrije Universiteit Amsterdam, De Boelelaan 1105, 1081 HV Amsterdam, The Netherlands}

% \author[4]{Manfred Jeusfeld}[%
% orcid=0000-0002-9421-8566,
% email=Manfred.Jeusfeld@acm.org,
% url=http://conceptbase.sourceforge.net/mjf/,
% \address[4]{University of Skövde, Högskolevägen 1, 541 28 Skövde, Sweden}

% \cortext[1]{Yi-Chun Chen}

% \fntext[1]{These authors contributed equally.}

\begin{abstract}
Understanding visual narratives such as comics requires structured representations that capture events, characters, and their relations across multiple levels of story organization. However, symbolic narrative graphs often suffer from inconsistency and redundancy, where similar actions or events are labeled differently across annotations or contexts. Such variance limits the effectiveness of reasoning and generalization.

This paper introduces a semantic normalization framework for hierarchical narrative knowledge graphs. Building on cognitively grounded models of narrative comprehension, we propose methods that consolidate semantically related actions and events using lexical similarity and embedding-based clustering. The normalization process reduces annotation noise, aligns symbolic categories across narrative levels, and preserves interpretability. 

We demonstrate the framework on annotated manga stories from the Manga109 dataset, applying normalization to panel-, event-, and story-level graphs. Preliminary evaluations across narrative reasoning tasks, such as action retrieval, character grounding, and event summarization, show that semantic normalization improves coherence and robustness, while maintaining symbolic transparency. These findings suggest that normalization is a key step toward scalable, cognitively inspired graph models for multimodal narrative understanding.

\end{abstract}


\maketitle

\section{Introduction}

\input{1_Intro}

\section{Related Work}

\input{2_Related}
\section{Methodology}

\input{3_Methodology}
\section{Experiments and Evaluation}

\input{4_Experiment}
\section{Results}
\input{5_Result}
\section{Discussion and Future Work}

\input{6_Discussion}

\section{Conclusion}

\input{8_Conclusion}
%%
%% Define the bibliography file to be used
\bibliography{sample-ceur}

\appendix


% All resources related to this study, including code, annotations, and examples, are available at the following GitHub repository:

% \begin{itemize}
%   \item \href{https://github.com/RimiChen/2025_VN_Semantic_Normalization}{\texttt{github.com/RimiChen/2025\_VN\_Semantic\_Normalization}}
% \end{itemize}

\section{Online Resources}

To support reproducibility and future research, all resources related to this study, including code, annotations, and graph construction tools, are open-sourced:

% Uncomment and insert link after acceptance:
\begin{itemize}
  \item \href{https://github.com/RimiChen/2025_VN_Semantic_Normalization}{\texttt{github.com/RimiChen/2025\_VN\_Semantic\_Normalization}}
\end{itemize}

\end{document}

%% file: 1_Intro.tex
Visual narratives, such as comics and graphic novels, integrate sequential images, textual dialogue, and spatial composition to communicate structured stories~\cite{mccloud1993understanding,cohn2013visual}. For artificial intelligence systems to interpret such media, it is necessary to capture not only the content of individual panels but also the evolving relationships among events across time. This requires structured representations that reflect multiple levels of abstraction, from fine-grained panel actions to broader narrative arcs.  

Cognitive theories of narrative comprehension suggest that readers segment visual sequences into hierarchical structures of semantically coherent events. For example, Cohn's Visual Narrative Grammar (VNG)~\cite{cohn2013visual} posits functional categories such as Establisher, Initial, Peak, and Release, which recursively combine into higher-order story units. Such theories highlight the importance of hierarchical organization for narrative understanding and motivate graph-based approaches that model panels, events, and macro-level arcs within a unified representation.  

Building on this foundation, prior work introduced a hierarchical knowledge graph framework for modeling narrative content across multiple abstraction levels. This representation integrates multimodal panel information with temporal and semantic relations at event and macro-event levels, enabling interpretable symbolic reasoning across the narrative hierarchy.  

In this paper, we extend that framework with a \textbf{semantic normalization module} designed to address a key limitation of symbolic systems: \textbf{annotation variability and redundancy}. Semantically similar actions or events are often labeled differently depending on context or annotator decisions, reducing robustness and limiting scalability across diverse narrative domains. The normalization module clusters related actions and events using lexical and embedding-based similarity, consolidating semantically redundant labels while preserving symbolic interpretability. By standardizing categories across contexts, normalization improves coherence and enables more reliable reasoning over multimodal narrative content.  

We evaluate the extended framework on annotated manga stories from the Manga109 dataset, focusing on narrative reasoning tasks such as semantic action retrieval, dialogue tracing, character grounding, timeline reconstruction, and event summarization. Our experiments show that normalization improves consistency in tasks affected by lexical variation while retaining the interpretability and modularity of symbolic graphs.  

\textbf{The contributions of this work are as follows:}
\begin{itemize}
    \item We identify annotation variability as a key barrier to scalable symbolic reasoning in visual narratives.  
    \item We extend an existing hierarchical narrative graph framework with a semantic normalization mechanism that consolidates related narrative actions and events while maintaining symbolic transparency.  
    \item We demonstrate that normalization enhances coherence and robustness across multiple narrative reasoning tasks on the Manga109 dataset.  
\end{itemize}  

This work advances cognitively inspired AI for narrative understanding by showing how semantic normalization can strengthen symbolic frameworks, bridging the gap between interpretability and robustness. The extended framework provides a foundation for downstream applications such as story retrieval, narrative analysis, and human-AI co-creative systems.  

\begin{figure*}[htbp]
    \centering
    \includegraphics[width=0.9\textwidth]{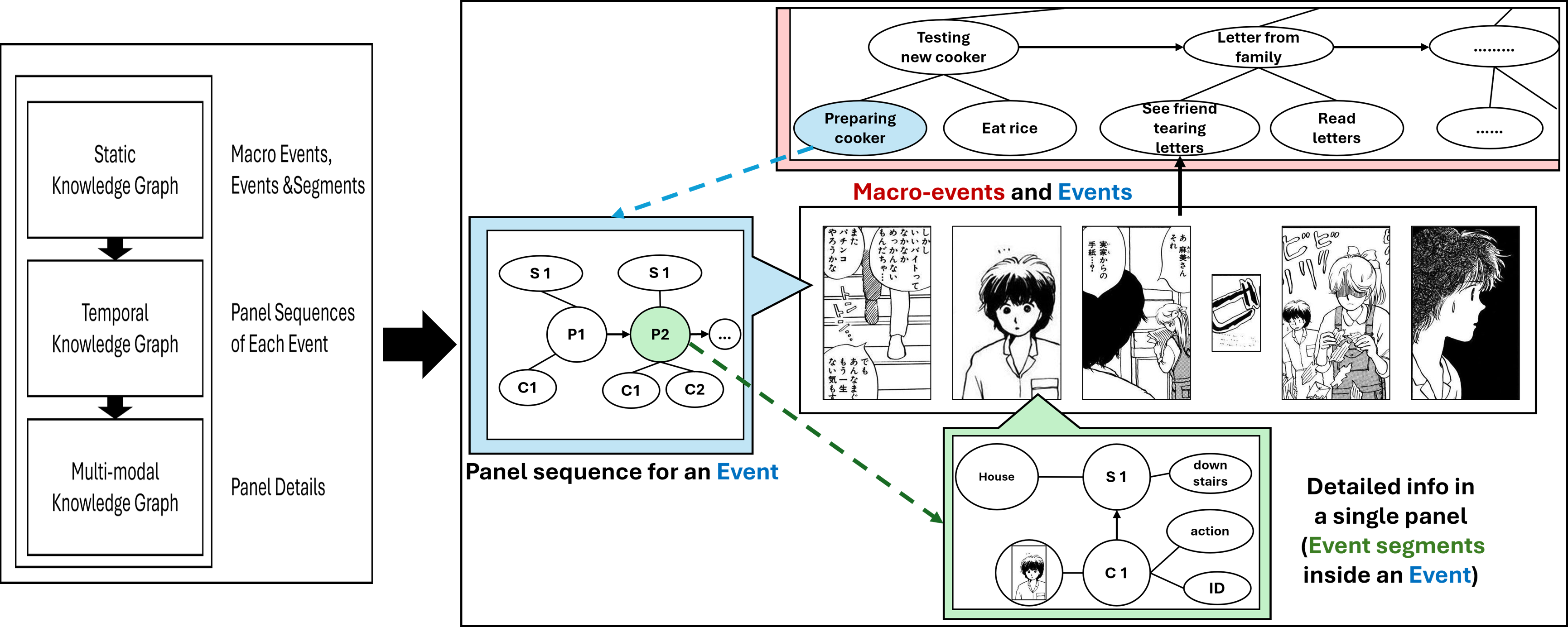}
    \caption{Visual narratives exhibit a hierarchical structure consisting of story arcs, events, and fine-grained panels. Each level contributes semantic and temporal information, while panels convey multimodal cues through image, text, and layout. The hierarchical graph framework integrates these levels into a unified representation, and our semantic normalization extension improves robustness by consolidating redundant action and event labels.}
    \label{fig:basic_structure}
\end{figure*}

%% file: 2_Related.tex
Narrative modeling and visual storytelling have been studied across cognitive science, artificial intelligence, and multimodal learning. This section reviews related work in four areas: (1) narrative event segmentation, (2) knowledge graphs for narrative reasoning, (3) semantic normalization in symbolic reasoning, and (4) narrative question answering and event-based inference. Together, these domains motivate the hierarchical knowledge graph framework introduced in prior work and the semantic normalization extension evaluated here.  

\subsection{Narrative Modeling and Event Segmentation}

Cognitive research has long examined how humans segment and comprehend stories. Theories of story grammar~\cite{thorndyke1977cognitive} and situation models~\cite{zwaan1995dimensions, zwaan1996processing, kelter2004representing} describe how shifts in time, space, or characters anchor comprehension and memory. Event segmentation theory~\cite{zacks2007event} extends this view by treating such shifts as psychologically meaningful boundaries that structure narrative understanding~\cite{kurby2008segmentation, sargent2013event}.  

These insights have inspired computational models that attempt to capture hierarchical segmentation, from biologically motivated cortical reservoir systems~\cite{dominey2021narrative} to large language model–based structure learning~\cite{michelmann2025large}. Work on visual narratives, particularly in manga, has explored segmentation using statistical and rule-based methods~\cite{chen2021computational, chen2023framework}, though these approaches often lack symbolic structures capable of supporting reasoning across multiple abstraction levels.  

Theories such as Visual Narrative Grammar (VNG)~\cite{cohn2013visual} emphasize that panels serve functional roles (e.g., Establisher, Initial, Peak, Release) which are recursively embedded into larger units. This view aligns with event segmentation theory and underscores the need for hierarchical representations that move beyond flat or sequential encodings.  

\subsection{Knowledge Graphs for Multimodal Narrative Reasoning}

Knowledge graphs (KGs) provide structured modeling tools that have been applied to diverse narrative tasks, including event linking, character modeling, and semantic grounding~\cite{liang2024survey, lymperaiou2024survey}. They have been used in story generation from data~\cite{mishra2019storytelling}, media synthesis with linked data~\cite{renzi2023storytelling}, and hierarchical plot modeling~\cite{akimoto2017computational, lee2020learning}. More recent efforts integrate visual signals into KG-based frameworks, for example by grounding characters and actions in multimodal graphs~\cite{hsu2020knowledge, xu2021imagine}.  

Despite these advances, most approaches rely on flat or sequence-aligned structures, limiting their ability to capture compositional relationships across modalities. Recent work introduced multi-level symbolic graphs for visual narratives~\cite{chen2025structured}, demonstrating the potential of hierarchical graph abstractions. The present study builds on that direction by evaluating an extension with semantic normalization, focusing on how it improves coherence and generalization across varied narrative types.  

\subsection{Semantic Normalization and Symbolic Reasoning}

Symbolic reasoning frameworks often face challenges with linguistic variability in free-text annotations, where semantically similar actions or events are inconsistently labeled. Normalization techniques that cluster or canonicalize labels using lexical or embedding-based similarity offer a practical solution. Such approaches build on studies of semantic similarity~\cite{mihalcea2006corpus, vulic2020probing} and connect with broader work in neuro-symbolic systems that integrate structured logic with learned representations~\cite{amizadeh2020neuro, yi2018neural, li2021calibrating}.  

Within multimodal contexts, normalization serves as a lightweight yet effective mechanism for aligning symbolic categories while retaining interpretability. Surveys of neuro-symbolic methods in vision~\cite{khan2025survey} suggest that such integration can bridge symbolic transparency and neural flexibility. This study adopts that perspective by incorporating normalization into hierarchical narrative graphs as a means of improving symbolic generalization.  

\subsection{Narrative Question Answering and Event-Based Inference}

Narrative question answering benchmarks highlight the role of event structure and temporal reasoning. Resources such as the Story Cloze Test~\cite{mostafazadeh2016corpus}, NarrativeQA~\cite{kovcisky2018narrativeqa}, and FairytaleQA~\cite{xu2022fantastic} focus on text-based narratives, while multimodal datasets like DramaQA~\cite{choi2021dramaqa} and MovieQA~\cite{tapaswi2016movieqa} extend the problem to video and character-grounded reasoning.  

Although most QA systems rely on pretrained neural encoders, symbolic structures have been shown to improve interpretability and coherence in story modeling~\cite{martin2018event}. The current study follows this line by examining hierarchical graphs spanning panel-, event-, and story-level abstractions, and by testing how semantic normalization influences reasoning performance across multimodal narratives.  

\paragraph{Summary.}  
The reviewed literature emphasizes the need for hierarchical, multimodal, and semantically adaptable structures in narrative understanding. Existing approaches either prioritize interpretability but lack flexibility, or rely on neural encoders while neglecting symbolic structure. The present work contributes to this landscape by extending a previously introduced hierarchical knowledge graph framework with semantic normalization, enabling more robust yet interpretable reasoning across narrative abstraction layers.  

%% file: 3_Methodology.tex
This study builds on a previously introduced hierarchical knowledge graph framework for modeling visual narratives~\cite{chen2025structured}. The original framework captures multimodal content, temporal progression, and semantic structure through three graph layers: panel-level multimodal graphs, sequence-level temporal graphs, and event-level semantic graphs. In this work, we retain that multi-layered design and extend it with a semantic normalization module that improves robustness by addressing variability in annotation labels. All components are implemented on a manually annotated subset of the Manga109 dataset.

\subsection{Dataset and Annotation Schema}

We use a curated subset of the Manga109 dataset~\cite{fujimoto2016manga109}, which contains professionally published Japanese comics across diverse genres. Each page is segmented into panels and annotated with perceptual and narrative elements, including character actions, visual objects, dialogue, and narrative roles.  

Annotations are organized into a three-tier hierarchy:  
\begin{itemize}
    \item \textbf{Panel-level:} fine-grained descriptions of actions, objects, and dialogue.  
    \item \textbf{Event-level:} mid-level narrative units composed of multiple panels.  
    \item \textbf{Macro-event level:} broad arcs spanning multiple events, marking thematic or structural shifts.  
\end{itemize}

This schema supports constructing graphs that represent grounded panel content, sequential flow, and abstract story arcs.

\begin{figure}[t]
  \centering
  \includegraphics[width=0.6\linewidth]{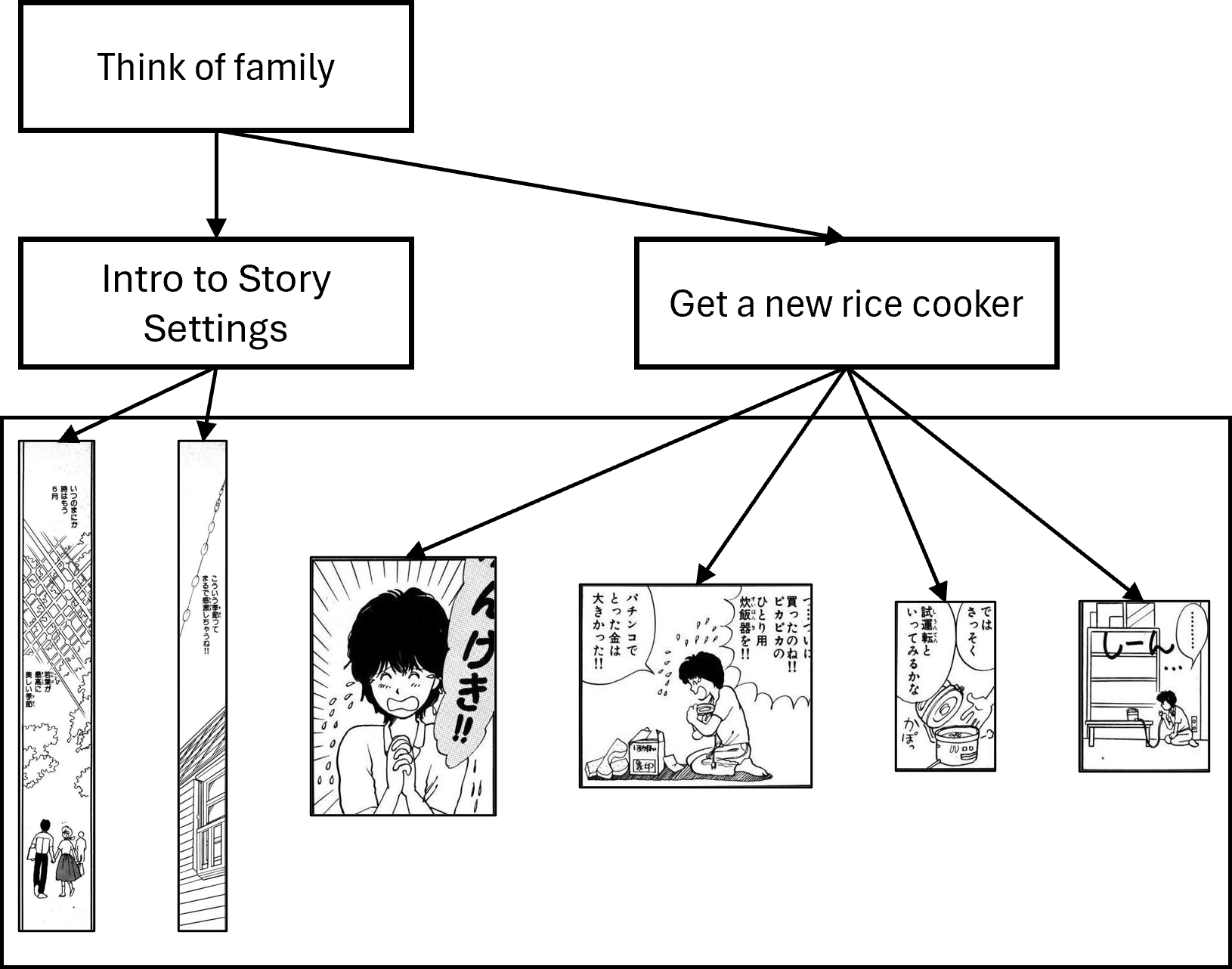}
  \caption{Hierarchical narrative event design. Panels are grouped into event segments, which instantiate mid-level events and aggregate into macro-events.}
  \label{fig:event-hierarchy}
\end{figure}

\subsection{Hierarchical Knowledge Graph Framework}

The baseline representation integrates three graph layers:

\paragraph{Panel-Level Multimodal Graphs.}  
Nodes capture characters, objects, actions, and dialogue, with edges encoding co-occurrence, agent-action relations, and text-image links.  

\paragraph{Sequence-Level Temporal Graphs.}  
Panels and event segments are linked with directed edges such as \textit{precedes\_reading} and \textit{precedes\_storytime}, supporting reasoning about narrative order.  

\paragraph{Event-Level Semantic Graphs.}  
Events and macro-events are connected with \textit{subevent-of}, \textit{precedes}, and \textit{co-occurs} edges, supporting abstract reasoning such as summarization.  

These layers are linked through \textit{instantiates} and \textit{subevent-of} relations, enabling reasoning both top-down and bottom-up.

\subsection{Semantic Normalization Module}

The primary methodological extension is a semantic normalization module. Symbolic reasoning can be sensitive to lexical variability in annotation labels. For example, actions annotated as ``attack,'' ``strike,'' or ``fight'' may refer to the same underlying concept. The normalization module consolidates such cases while preserving symbolic interpretability.

\paragraph{Input.}  
Action and event labels from panel and event annotations.

\paragraph{Lexical Normalization.}  
Lemmatization and synonym expansion (via WordNet) reduce inflectional and surface variation.

\paragraph{Embedding-Based Similarity.}  
Sentence-level embeddings (Sentence-BERT) are used to compute cosine similarity among labels. Labels exceeding a similarity threshold are grouped.

\paragraph{Clustering and Relabeling.}  
A graph is built in the similarity space, with connected components forming equivalence clusters. Each cluster is assigned a canonical label using a heuristic: gold annotation labels are preferred; otherwise, the shortest surface form is selected.

\paragraph{Integration.}  
Normalization is applied after graph construction and before downstream reasoning. Tasks such as action retrieval, dialogue tracing, and event summarization operate on the normalized graph.

This process adapts canonicalization techniques from ontology alignment~\cite{shvaiko2011ontology}, entity normalization in knowledge bases~\cite{lehmann2015dbpedia}, and QA dataset normalization~\cite{rajpurkar2016squad}.

\paragraph{Illustrative Example.}  
In a combat scene, labels such as ``attacks,'' ``strikes,'' and ``fights'' are grouped into one canonical category. This reduces annotation sparsity and improves the reliability of symbolic reasoning.

\subsection{Implementation Details}

Graph construction is implemented in Python with \texttt{NetworkX}. Both raw and normalized graph variants are serialized in JSON for flexibility in downstream use. Code, schemas, and sample annotations will be released publicly upon acceptance.

%% file: 4_Experiment.tex
\label{sec:evaluation}

We conduct experiments to evaluate the effect of semantic normalization within the hierarchical knowledge graph framework. The baseline framework~\cite{chen2025structured} provides structured representations across panel, sequence, and event levels; here we extend it with normalization and assess its contribution to symbolic reasoning under conditions of lexical variability. We adapt the evaluation tasks from prior work and re-assess them with and without normalization to probe multimodal grounding, temporal coherence, and hierarchical abstraction.

\subsection{Dataset and Graph Construction}
\label{sec:exp-setup}

Experiments use a manually annotated subset of the Manga109 dataset~\cite{fujimoto2016manga109}, which includes professionally published Japanese comics across diverse genres and artistic styles. Each page is segmented into panels and annotated with multimodal content, including characters, actions, objects, dialogue, captions, camera framing, and hierarchical narrative labels.  

Two graph variants are evaluated:  
(1) \textbf{Raw graphs} without normalization, and  
(2) \textbf{Normalized graphs} in which action and event labels are standardized.  

\subsection{Reasoning Tasks}

We evaluate five tasks that reflect essential aspects of narrative comprehension:

\paragraph{1. Semantic Action Retrieval.}  
Given a canonical action label (e.g., \textit{attack}), the system retrieves all panel segments expressing semantically related actions (e.g., \textit{strike}, \textit{fight}). This task is directly affected by normalization.  

\paragraph{2. Dialogue Tracing.}  
For a given event, the system identifies associated dialogue spans and speaker entities across its panels, testing cross-modal grounding.  

\paragraph{3. Character Trajectory Mapping.}  
The system traces a character’s appearances across panels and events, evaluating identity continuity across narrative levels.  

\paragraph{4. Timeline Reconstruction.}  
Temporal orderings are recovered by traversing \textit{precedes\_reading} and \textit{precedes\_storytime} edges.  

\paragraph{5. Event Summarization.}  
Constituent sub-events or panel segments are retrieved for a given event or macro-event, testing support for hierarchical decomposition.  

\subsection{Evaluation Protocol}

Both qualitative and quantitative evaluation are performed:  

\begin{itemize}
    \item \textbf{Qualitative Evaluation:} Manual inspection assesses interpretability, symbolic alignment, and narrative coherence across tasks.  
    \item \textbf{Quantitative Metrics:}
    \begin{itemize}
        \item \textbf{Action Retrieval F1:} Precision/recall of retrieved actions, with and without normalization.  
        \item \textbf{Dialogue Trace F1:} Token-level overlap between predicted and annotated dialogue spans.  
        \item \textbf{Trajectory Coverage:} Proportion of annotated character appearances retrieved successfully.  
        \item \textbf{Ordering Accuracy:} Agreement between predicted and annotated narrative sequences.  
        \item \textbf{Event Summary F1:} Overlap between retrieved and annotated sub-events.  
    \end{itemize}
\end{itemize}

\noindent\textit{Note:} Development-set examples are included to illustrate reasoning behavior. Final quantitative results are reported in Section~\ref{sec:results}.

\subsection{Development Set Examples}

Representative examples highlight reasoning supported by the graph framework and improved by normalization:

\paragraph{Action Retrieval.}  
Querying \textit{attack} retrieves \textit{fight}, \textit{strike}, and \textit{hit}, illustrating normalization benefits for aligning related actions.  

\paragraph{Dialogue Tracing (\textit{Monster intro}).}  
The system associates four dialogue segments with three panels and two speakers, showing effective text–image linking.  

\paragraph{Character Trajectory (Character A).}  
The character appears in twelve panels across four events and two macro-events. Identity continuity is preserved via \textit{refers-to} and \textit{has-agent} edges.  

\paragraph{Event Summarization (\textit{Think of family}).}  
The retrieved sub-event sequence (\textit{Intro}, \textit{Get new rice cooker}, \textit{Test new rice cooker}, \textit{Eat and think of family}) matches the annotated structure.  

\paragraph{Timeline Reconstruction (Macro-event 1).}  
The predicted panel order \texttt{[0\_0\_0, 0\_0\_1, ..., 0\_2\_3]} corresponds to the annotated reading sequence, demonstrating accurate traversal of temporal edges.  

\subsection{Ground Truth Construction}

Gold references are derived from manual annotations:  

\begin{itemize}
    \item \textbf{Action Retrieval:} Canonical clusters defined by normalization.  
    \item \textbf{Dialogue Tracing:} Dialogue spans and speaker links annotated per panel.  
    \item \textbf{Character Trajectory:} Character appearances tagged per panel, linked by identity edges.  
    \item \textbf{Timeline Reconstruction:} Temporal orderings annotated at panel and event levels.  
    \item \textbf{Event Summarization:} Event hierarchies defined by \textit{subevent-of} relations.  
\end{itemize}

%% file: 5_Result.tex
\label{sec:evaluation}

\label{sec:results}

We evaluate the hierarchical knowledge graph framework, extended with a semantic normalization module, across two contrasting manga stories: \textit{Aisazu Niha Irarenai} (romance) and \textit{Akkera Kanjinchou} (battle). These genres were chosen to assess how symbolic reasoning generalizes across subtle, introspective storylines and fast-paced, visually dynamic narratives.  

Each story is evaluated on five reasoning tasks: (1) semantic action retrieval, (2) dialogue tracing, (3) character trajectory mapping, (4) timeline reconstruction, and (5) event summarization. We report F1 scores for Tasks 1, 2, 3, and 5, while Task 4 achieves near-perfect accuracy and is summarized only. Results are presented for both raw graphs and semantically normalized graphs to highlight the effect of normalization.  

\subsection{Results on \textit{Aisazu Niha Irarenai} (Romance Genre)}

Table~\ref{tab:results_s0} summarizes performance across macro-events. Tasks 2 (dialogue tracing), 3 (character trajectory), and 4 (timeline reconstruction) all achieve perfect scores, confirming the reliability of the baseline framework for maintaining dialogue attribution, character identity, and event order.  

Task 1 (semantic action retrieval) shows slight degradation after normalization, largely due to mismatches between normalized and canonical labels (e.g., \texttt{insert} vs. \texttt{insert\_into}). Task 5 (event summarization) yields moderate scores, reflecting the inherent difficulty of segmenting narrative abstraction boundaries in introspective storylines.  

\begin{table}[t]
\caption{Reasoning performance on \textit{Aisazu Niha Irarenai} (romance genre). Task 4 (timeline reconstruction) achieved F1 = 1.00 and is omitted.}
\label{tab:results_s0}
\centering
\small
\resizebox{\columnwidth}{!}{
\begin{tabular}{|l|c|c|c|c|c|}
\hline
\textbf{Macro-event} & \textbf{T1 F1 (raw)} & \textbf{T1 F1 (norm)} & \textbf{T2 F1} & \textbf{T3 F1} & \textbf{T5 F1} \\\hline
Message from family & 0.97 & 0.93 & 1.00 & 1.00 & 0.43 \\
Shock by message & 1.00 & 1.00 & 1.00 & 1.00 & 0.57 \\
Think of family & 0.92 & 0.78 & 1.00 & 1.00 & 0.62 \\
\hline
\end{tabular}
}
\end{table}

\subsection{Results on \textit{Akkera Kanjinchou} (Battle Genre)}

Table~\ref{tab:results_s1} reports performance on the battle-genre story. Task 1 achieves perfect or near-perfect scores, with little effect from normalization since the underlying annotations were already consistent. Task 2 (dialogue tracing) shows modest variation, with occasional mismatches in span segmentation. Tasks 3 and 4 again reach perfect or near-perfect accuracy, confirming robust identity and temporal tracking.  

Task 5 highlights challenges in summarizing abstract event structure within visually dense, fast-paced narratives. While most sub-event relations are correctly recovered, boundary detection occasionally over-clusters adjacent sequences or overlooks subtle semantic transitions.  

\begin{table}[t]
\caption{Reasoning performance on \textit{Akkera Kanjinchou} (battle genre). Task 4 (timeline reconstruction) achieved F1 $\geq$ 0.91 and is omitted for brevity.}
\label{tab:results_s1}
\centering
\small
\resizebox{\columnwidth}{!}{
\begin{tabular}{|l|c|c|c|c|c|}
\hline
\textbf{Macro-event} & \textbf{T1 F1 (raw)} & \textbf{T1 F1 (norm)} & \textbf{T2 F1} & \textbf{T3 F1} & \textbf{T5 F1} \\\hline
Intro to timeline & 1.00 & 1.00 & 0.50 & 1.00 & 0.67 \\
Intro main character & 0.80 & 0.80 & 1.00 & 1.00 & 0.50 \\
Intro second character & 1.00 & 1.00 & 0.91 & 1.00 & 0.50 \\
Monster intro & 1.00 & 1.00 & 0.80 & 1.00 & 0.33 \\
Street accident & 1.00 & 1.00 & 1.00 & 1.00 & 0.50 \\
Meet characters & 1.00 & 1.00 & 1.00 & 1.00 & 0.33 \\
Formal monster intro & 1.00 & 1.00 & 1.00 & 1.00 & 0.50 \\
\hline
\end{tabular}
}
\end{table}

\begin{figure}[t]
  \centering
  \begin{subfigure}[t]{0.49\textwidth}
    \centering
    \includegraphics[width=\linewidth]{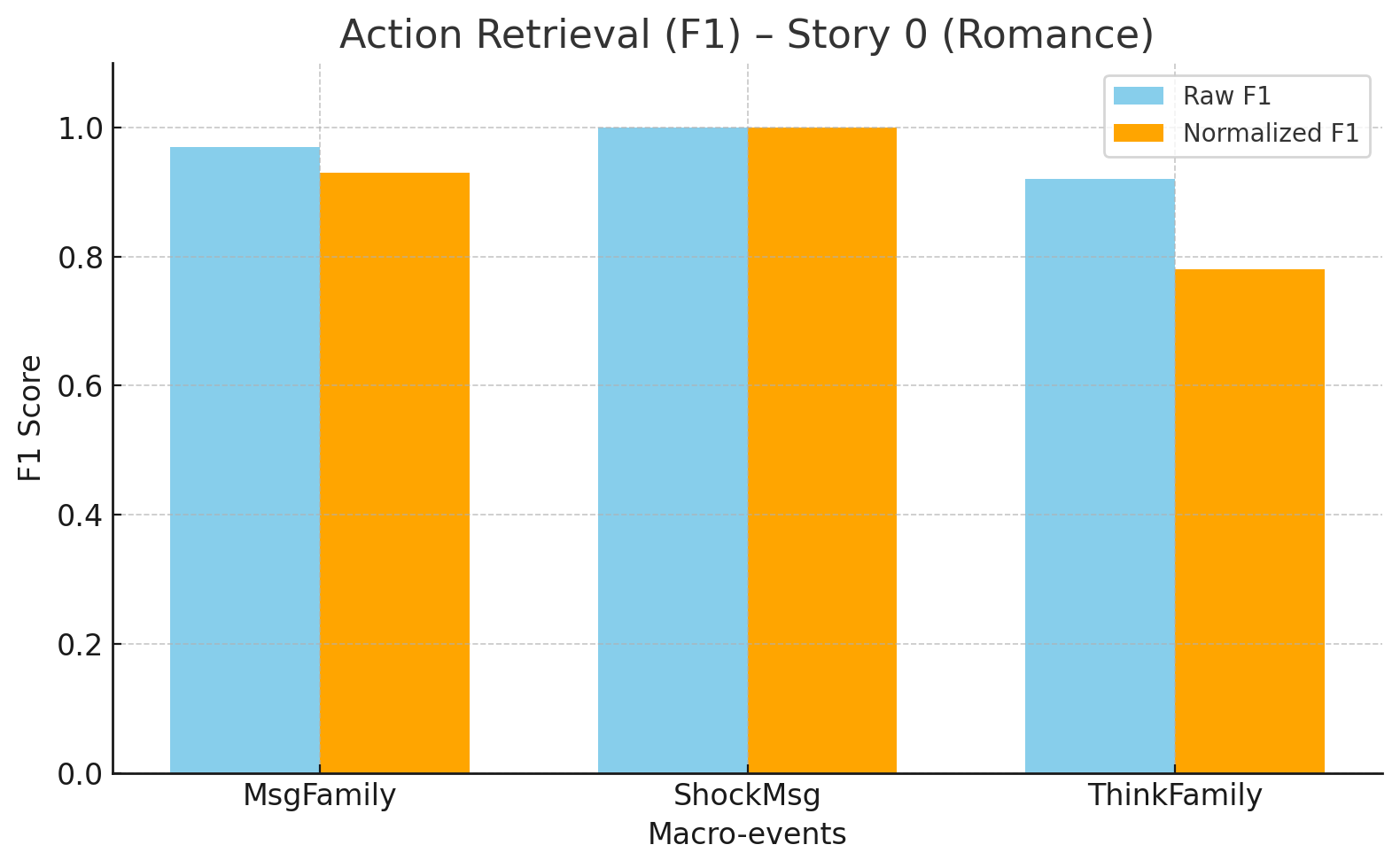}
    \caption{\textit{Aisazu Niha Irarenai} (romance). Normalization introduces both improvements and degradations depending on label alignment.}
    \label{fig:action-story0}
  \end{subfigure}
  \hfill
  \begin{subfigure}[t]{0.49\textwidth}
    \centering
    \includegraphics[width=\linewidth]{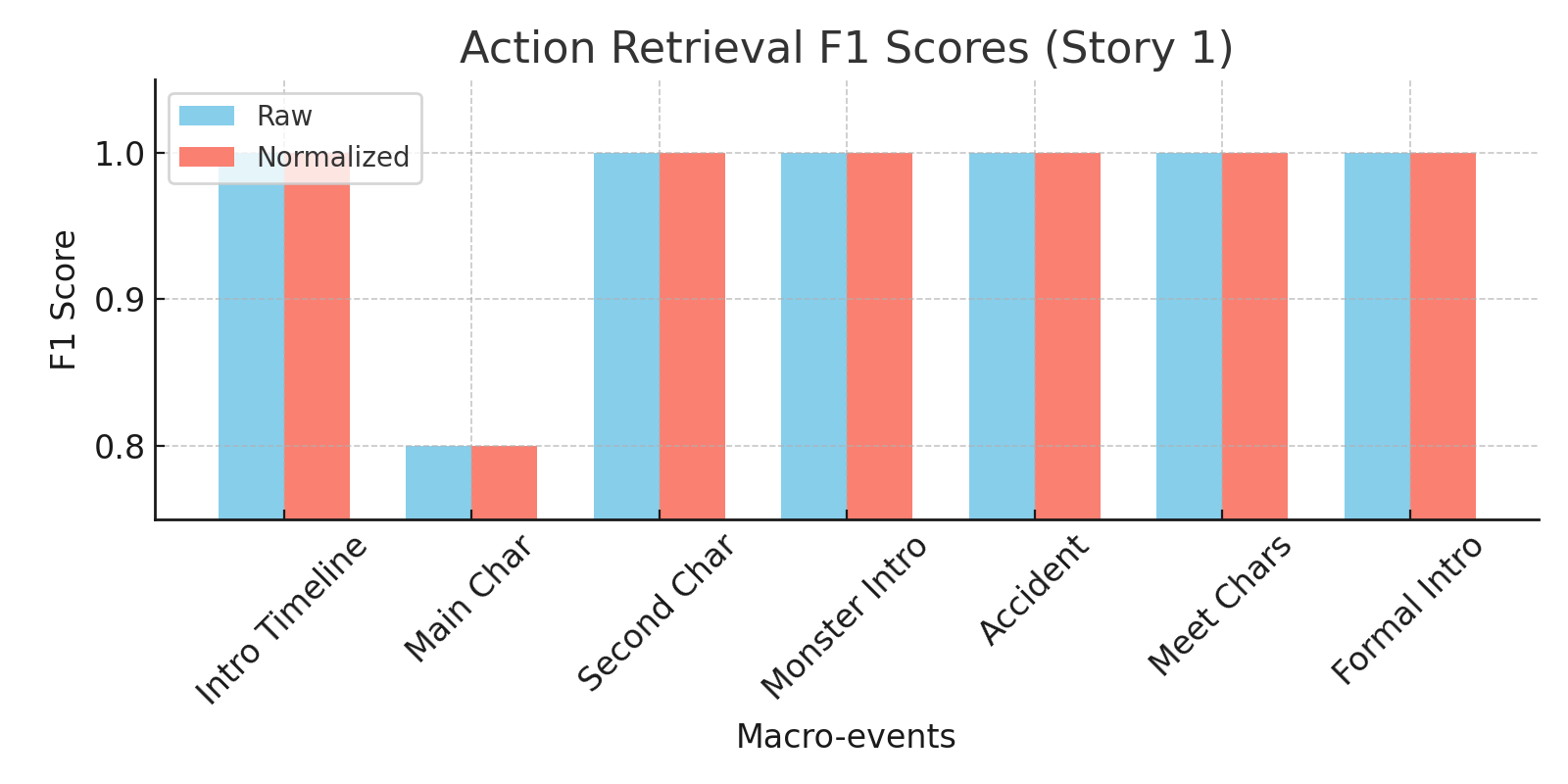}
    \caption{\textit{Akkera Kanjinchou} (battle). Consistent annotations reduce the effect of normalization.}
    \label{fig:action-story1}
  \end{subfigure}
  \caption{Macro-event--level F1 scores for semantic action retrieval (Task 1) across two genres.}
  \label{fig:action-story-combined}
\end{figure}

\subsection{Summary and Insights}

The results confirm that the hierarchical knowledge graph framework provides a stable basis for symbolic reasoning, consistently achieving high accuracy on tasks involving character tracking, temporal order, and dialogue recovery.  

The semantic normalization extension shows selective but meaningful benefits. In the romance genre, where action labels are lexically diverse, normalization improved retrieval consistency but also introduced occasional mismatches when canonicalization diverged from annotations. In the battle genre, where annotations were already consistent, normalization had little measurable effect.  

Overall, normalization is most useful in narrative domains with high linguistic variability, while the baseline symbolic framework reliably supports reasoning across narrative levels. This underscores the value of normalization as an incremental but effective extension rather than a universal improvement.

%% file: 6_Discussion.tex
\label{sec:discussion}

\label{sec:discussion}

This section reflects on the experimental findings and design decisions underlying the hierarchical knowledge graph framework and its semantic normalization extension. We analyze performance trends across genres and reasoning tasks, examine the benefits and trade-offs of normalization, and highlight implications for symbolic narrative modeling. Together, these insights situate the extension within broader efforts in interpretable, cognitively grounded multimodal AI.

Experimental results reaffirm that the hierarchical knowledge graph framework reliably supports accurate symbolic reasoning across diverse narrative tasks. The added semantic normalization mitigates lexical variability in action and event labels, contributing to greater consistency when annotations are heterogeneous. At the same time, normalization occasionally introduced mismatches that reduced task performance, underscoring the trade-off between generalization and precision.

\subsection{Cross-Story Performance and Robustness}

The baseline framework performed consistently across contrasting genres with distinct stylistic conventions. In the romance story \textit{Aisazu Niha Irarenai}, dialogue tracing, character mapping, and timeline reconstruction all achieved perfect scores, while action retrieval showed sensitivity to label variation. Normalization improved retrieval in some cases but degraded it in others when canonical labels diverged from annotations. In the battle story \textit{Akkera Kanjinchou}, where action annotations were already consistent, normalization had little effect. Across both stories, event summarization remained the most challenging task, reflecting the inherent difficulty of segmenting abstract narrative hierarchies and aligning sub-event boundaries.

\subsection{Framework Strengths}

The extended framework preserves the strengths of the original design while adding selective improvements. Its interpretable, compositional architecture integrates multimodal panel-level content with temporal and semantic abstractions at higher levels, enabling reasoning across the full narrative hierarchy. Symbolic representations also allow for modular querying and controlled manipulation of reasoning paths—capabilities difficult to achieve with purely neural systems. These features make the framework particularly suitable for explainable AI, narrative analysis, and co-creative authoring environments.

\subsection{Limitations and Open Challenges}

Despite these strengths, several challenges remain. Performance in action retrieval was constrained by annotation variability, and normalization occasionally overgeneralized, reducing alignment with gold labels. Evaluation protocols relying on exact string matching further underrepresent conceptual similarity. Finally, the construction of hierarchical annotations remains costly. Table~\ref{tab:limitations} summarizes key limitations and potential mitigation strategies.

\begin{table}[t]
\caption{Summary of limitations and proposed mitigation strategies.}
\label{tab:limitations}
\centering
\small
\begin{tabular}{|p{6cm}|p{6cm}|}
\hline
\textbf{Limitation} & \textbf{Potential Mitigation} \\
\hline
Lexical variation in action labels leading to semantic mismatches & Task-specific normalization with curated canonical labels or lightweight verb ontologies \\
\hline
Overgeneralization in normalization misaligned with gold annotations & Human-in-the-loop correction or post-hoc refinement using ground truth labels \\
\hline
Evaluation using exact string matching underrepresents semantic similarity & Embedding-based or paraphrase-aware evaluation metrics \\
\hline
High manual cost for constructing multi-level structures & Semi-automated annotation using pretrained vision-language models \\
\hline
\end{tabular}
\end{table}

An illustrative case is the macro-event ``Think of family,'' where normalization reduced F1 due to mismatches such as \texttt{insert} vs. \texttt{insert\_into}, despite their semantic equivalence. Such examples highlight the need for evaluation methods that capture conceptual similarity rather than surface form alone.

\subsection{Future Directions}

Future work will focus on improving both generalizability and usability. Planned directions include developing task-specific normalization schemes, incorporating lightweight semantic ontologies for more robust label alignment, and expanding evaluation protocols to include semantic-aware metrics. Semi-automated annotation pipelines leveraging pretrained vision-language models will also be explored to reduce manual effort in constructing hierarchical graphs.

In the longer term, the framework is expected to support co-creative and generative applications, including branching story generation, interactive narrative systems, and author-guided content editing. Symbolic abstractions provide a flexible and interpretable scaffold for guiding layout, event planning, and dialogue coherence. Prior work on customizable visual storytelling~\cite{chen2023customizable,chen2024collaborative} suggests that such structured representations can enhance control and creative agency in multimodal generation.

\paragraph{Summary.}  
This study evaluates an incremental extension to hierarchical narrative graph modeling through the incorporation of semantic normalization. The results highlight both its benefits in reducing lexical variability and its trade-offs in maintaining alignment with gold annotations. Together, these findings reinforce the potential of symbolic graph structures as scaffolds for interpretable reasoning and as a foundation for future multimodal narrative AI.

%% file: 8_Conclusion.tex
\label{sec:conclusion}

This work extends an existing hierarchical knowledge graph framework for visual narrative comprehension by incorporating a semantic normalization module. The base framework organizes narrative content at three levels—panel, event, and macro-event—capturing temporal, semantic, and multimodal relations in an integrated symbolic structure. The normalization component reduces lexical variability in action and event labels, fostering greater consistency while preserving symbolic interpretability.  

We evaluated the extended framework on five reasoning tasks: action retrieval, dialogue tracing, character trajectory mapping, timeline reconstruction, and event summarization. Experiments on two manga stories from contrasting genres (romance and battle) showed that normalization yielded moderate gains in contexts with higher lexical variation, while its benefits were limited when annotations were already consistent. These findings clarify both the strengths and trade-offs of normalization in symbolic narrative modeling.  

Beyond task-level outcomes, the results reaffirm that symbolic and hierarchical representations provide a reliable basis for interpretable reasoning across multiple abstraction levels. The framework’s modular design supports applications in narrative analysis, explainable AI, and human-in-the-loop authoring, while the normalization layer demonstrates a pathway to improving robustness without undermining transparency.  

Future work will pursue three directions: (1) refining normalization with task-specific schemes, embedding-based similarity measures, and lightweight ontologies; (2) scaling to larger and more diverse narrative corpora through semi-automated annotation; and (3) adapting the representation to additional narrative domains such as animation, visual novels, and interactive fiction.  

\paragraph{Summary.}  
This study contributes an incremental extension to hierarchical narrative graph modeling. By integrating semantic normalization, it strengthens symbolic reasoning under variable annotation conditions and offers insight into the broader challenge of balancing lexical flexibility with structured interpretability in multimodal narrative AI.

%% file: 0_AIC_Main.bbl
\begin{thebibliography}{40}
\expandafter\ifx\csname natexlab\endcsname\relax\def\natexlab#1{#1}\fi
\providecommand{\url}[1]{\texttt{#1}}
\providecommand{\href}[2]{#2}
\providecommand{\path}[1]{#1}
\providecommand{\DOIprefix}{doi:}
\providecommand{\ArXivprefix}{arXiv:}
\providecommand{\URLprefix}{URL: }
\providecommand{\Pubmedprefix}{pmid:}
\providecommand{\doi}[1]{\href{http://dx.doi.org/#1}{\path{#1}}}
\providecommand{\Pubmed}[1]{\href{pmid:#1}{\path{#1}}}
\providecommand{\bibinfo}[2]{#2}
\ifx\xfnm\relax \def\xfnm[#1]{\unskip,\space#1}\fi
%Type = Book
\bibitem[{McCloud(1993)}]{mccloud1993understanding}
\bibinfo{author}{S.~McCloud}, \bibinfo{title}{Understanding Comics: The Invisible Art}, \bibinfo{publisher}{Tundra Publishing}, \bibinfo{year}{1993}. \bibinfo{note}{Reprinted by HarperCollins in 1994}.
%Type = Article
\bibitem[{Cohn(2013)}]{cohn2013visual}
\bibinfo{author}{N.~Cohn},
\newblock \bibinfo{title}{Visual narrative structure},
\newblock \bibinfo{journal}{Cognitive science} \bibinfo{volume}{37} (\bibinfo{year}{2013}) \bibinfo{pages}{413--452}.
%Type = Article
\bibitem[{Thorndyke(1977)}]{thorndyke1977cognitive}
\bibinfo{author}{P.~W. Thorndyke},
\newblock \bibinfo{title}{Cognitive structures in comprehension and memory of narrative discourse},
\newblock \bibinfo{journal}{Cognitive psychology} \bibinfo{volume}{9} (\bibinfo{year}{1977}) \bibinfo{pages}{77--110}.
%Type = Article
\bibitem[{Zwaan et~al.(1995)Zwaan, Magliano, and Graesser}]{zwaan1995dimensions}
\bibinfo{author}{R.~A. Zwaan}, \bibinfo{author}{J.~P. Magliano}, \bibinfo{author}{A.~C. Graesser},
\newblock \bibinfo{title}{Dimensions of situation model construction in narrative comprehension.},
\newblock \bibinfo{journal}{Journal of experimental psychology: Learning, memory, and cognition} \bibinfo{volume}{21} (\bibinfo{year}{1995}) \bibinfo{pages}{386}.
%Type = Article
\bibitem[{Zwaan(1996)}]{zwaan1996processing}
\bibinfo{author}{R.~A. Zwaan},
\newblock \bibinfo{title}{Processing narrative time shifts.},
\newblock \bibinfo{journal}{Journal of Experimental Psychology: Learning, memory, and cognition} \bibinfo{volume}{22} (\bibinfo{year}{1996}) \bibinfo{pages}{1196}.
%Type = Article
\bibitem[{Kelter et~al.(2004)Kelter, Kaup, and Claus}]{kelter2004representing}
\bibinfo{author}{S.~Kelter}, \bibinfo{author}{B.~Kaup}, \bibinfo{author}{B.~Claus},
\newblock \bibinfo{title}{Representing a described sequence of events: a dynamic view of narrative comprehension.},
\newblock \bibinfo{journal}{Journal of Experimental Psychology: Learning, Memory, and Cognition} \bibinfo{volume}{30} (\bibinfo{year}{2004}) \bibinfo{pages}{451}.
%Type = Article
\bibitem[{Zacks and Swallow(2007)}]{zacks2007event}
\bibinfo{author}{J.~M. Zacks}, \bibinfo{author}{K.~M. Swallow},
\newblock \bibinfo{title}{Event segmentation},
\newblock \bibinfo{journal}{Current directions in psychological science} \bibinfo{volume}{16} (\bibinfo{year}{2007}) \bibinfo{pages}{80--84}.
%Type = Article
\bibitem[{Kurby and Zacks(2008)}]{kurby2008segmentation}
\bibinfo{author}{C.~A. Kurby}, \bibinfo{author}{J.~M. Zacks},
\newblock \bibinfo{title}{Segmentation in the perception and memory of events},
\newblock \bibinfo{journal}{Trends in cognitive sciences} \bibinfo{volume}{12} (\bibinfo{year}{2008}) \bibinfo{pages}{72--79}.
%Type = Article
\bibitem[{Sargent et~al.(2013)Sargent, Zacks, Hambrick, Zacks, Kurby, Bailey, Eisenberg, and Beck}]{sargent2013event}
\bibinfo{author}{J.~Q. Sargent}, \bibinfo{author}{J.~M. Zacks}, \bibinfo{author}{D.~Z. Hambrick}, \bibinfo{author}{R.~T. Zacks}, \bibinfo{author}{C.~A. Kurby}, \bibinfo{author}{H.~R. Bailey}, \bibinfo{author}{M.~L. Eisenberg}, \bibinfo{author}{T.~M. Beck},
\newblock \bibinfo{title}{Event segmentation ability uniquely predicts event memory},
\newblock \bibinfo{journal}{Cognition} \bibinfo{volume}{129} (\bibinfo{year}{2013}) \bibinfo{pages}{241--255}.
%Type = Article
\bibitem[{Dominey(2021)}]{dominey2021narrative}
\bibinfo{author}{P.~F. Dominey},
\newblock \bibinfo{title}{Narrative event segmentation in the cortical reservoir},
\newblock \bibinfo{journal}{PLOS Computational Biology} \bibinfo{volume}{17} (\bibinfo{year}{2021}) \bibinfo{pages}{e1008993}.
%Type = Article
\bibitem[{Michelmann et~al.(2025)Michelmann, Kumar, Norman, and Toneva}]{michelmann2025large}
\bibinfo{author}{S.~Michelmann}, \bibinfo{author}{M.~Kumar}, \bibinfo{author}{K.~A. Norman}, \bibinfo{author}{M.~Toneva},
\newblock \bibinfo{title}{Large language models can segment narrative events similarly to humans},
\newblock \bibinfo{journal}{Behavior Research Methods} \bibinfo{volume}{57} (\bibinfo{year}{2025}) \bibinfo{pages}{1--13}.
%Type = Inproceedings
\bibitem[{Chen and Jhala(2021)}]{chen2021computational}
\bibinfo{author}{Y.-C. Chen}, \bibinfo{author}{A.~Jhala},
\newblock \bibinfo{title}{A computational model of comprehension in manga style visual narratives},
\newblock in: \bibinfo{booktitle}{Proceedings of the Annual Meeting of the Cognitive Science Society}, volume~\bibinfo{volume}{43}, \bibinfo{year}{2021}.
%Type = Book
\bibitem[{Chen(2023)}]{chen2023framework}
\bibinfo{author}{Y.-C. Chen}, \bibinfo{title}{A Framework with Hierarchical Models for Visual Narrative Sequence Comprehension and its Applications}, \bibinfo{publisher}{North Carolina State University}, \bibinfo{year}{2023}.
%Type = Article
\bibitem[{Liang et~al.(2024)Liang, Meng, Liu, Liu, Tu, Wang, Zhou, Liu, Sun, and He}]{liang2024survey}
\bibinfo{author}{K.~Liang}, \bibinfo{author}{L.~Meng}, \bibinfo{author}{M.~Liu}, \bibinfo{author}{Y.~Liu}, \bibinfo{author}{W.~Tu}, \bibinfo{author}{S.~Wang}, \bibinfo{author}{S.~Zhou}, \bibinfo{author}{X.~Liu}, \bibinfo{author}{F.~Sun}, \bibinfo{author}{K.~He},
\newblock \bibinfo{title}{A survey of knowledge graph reasoning on graph types: Static, dynamic, and multi-modal},
\newblock \bibinfo{journal}{IEEE Transactions on Pattern Analysis and Machine Intelligence}  (\bibinfo{year}{2024}).
%Type = Article
\bibitem[{Lymperaiou and Stamou(2024)}]{lymperaiou2024survey}
\bibinfo{author}{M.~Lymperaiou}, \bibinfo{author}{G.~Stamou},
\newblock \bibinfo{title}{A survey on knowledge-enhanced multimodal learning},
\newblock \bibinfo{journal}{Artificial Intelligence Review} \bibinfo{volume}{57} (\bibinfo{year}{2024}) \bibinfo{pages}{284}.
%Type = Inproceedings
\bibitem[{Mishra et~al.(2019)Mishra, Laha, Sankaranarayanan, Jain, and Krishnan}]{mishra2019storytelling}
\bibinfo{author}{A.~Mishra}, \bibinfo{author}{A.~Laha}, \bibinfo{author}{K.~Sankaranarayanan}, \bibinfo{author}{P.~Jain}, \bibinfo{author}{S.~Krishnan},
\newblock \bibinfo{title}{Storytelling from structured data and knowledge graphs: An nlg perspective},
\newblock in: \bibinfo{booktitle}{Proceedings of the 57th annual meeting of the association for computational linguistics: Tutorial Abstracts}, \bibinfo{year}{2019}, pp. \bibinfo{pages}{43--48}.
%Type = Article
\bibitem[{Renzi et~al.(2023)Renzi, Rinaldi, Russo, and Tommasino}]{renzi2023storytelling}
\bibinfo{author}{G.~Renzi}, \bibinfo{author}{A.~M. Rinaldi}, \bibinfo{author}{C.~Russo}, \bibinfo{author}{C.~Tommasino},
\newblock \bibinfo{title}{A storytelling framework based on multimedia knowledge graph using linked open data and deep neural networks},
\newblock \bibinfo{journal}{Multimedia Tools and Applications} \bibinfo{volume}{82} (\bibinfo{year}{2023}) \bibinfo{pages}{31625--31639}.
%Type = Article
\bibitem[{Akimoto(2017)}]{akimoto2017computational}
\bibinfo{author}{T.~Akimoto},
\newblock \bibinfo{title}{Computational modeling of narrative structure: A hierarchical graph model for multidimensional narrative structure},
\newblock \bibinfo{journal}{International Journal of Computational Linguistics Research} \bibinfo{volume}{8} (\bibinfo{year}{2017}) \bibinfo{pages}{92--108}.
%Type = Article
\bibitem[{Lee et~al.(2020)Lee, Jung, and Kim}]{lee2020learning}
\bibinfo{author}{O.-J. Lee}, \bibinfo{author}{J.~J. Jung}, \bibinfo{author}{J.-T. Kim},
\newblock \bibinfo{title}{Learning hierarchical representations of stories by using multi-layered structures in narrative multimedia},
\newblock \bibinfo{journal}{Sensors} \bibinfo{volume}{20} (\bibinfo{year}{2020}) \bibinfo{pages}{1978}.
%Type = Inproceedings
\bibitem[{Hsu et~al.(2020)Hsu, Chen, Hsu, Li, Lin, Huang, and Ku}]{hsu2020knowledge}
\bibinfo{author}{C.-C. Hsu}, \bibinfo{author}{Z.-Y. Chen}, \bibinfo{author}{C.-Y. Hsu}, \bibinfo{author}{C.-C. Li}, \bibinfo{author}{T.-Y. Lin}, \bibinfo{author}{T.-H. Huang}, \bibinfo{author}{L.-W. Ku},
\newblock \bibinfo{title}{Knowledge-enriched visual storytelling},
\newblock in: \bibinfo{booktitle}{Proceedings of the AAAI Conference on Artificial Intelligence}, volume~\bibinfo{volume}{34}, \bibinfo{year}{2020}, pp. \bibinfo{pages}{7952--7960}.
%Type = Inproceedings
\bibitem[{Xu et~al.(2021)Xu, Yang, Li, Shen, Ao, and Xu}]{xu2021imagine}
\bibinfo{author}{C.~Xu}, \bibinfo{author}{M.~Yang}, \bibinfo{author}{C.~Li}, \bibinfo{author}{Y.~Shen}, \bibinfo{author}{X.~Ao}, \bibinfo{author}{R.~Xu},
\newblock \bibinfo{title}{Imagine, reason and write: Visual storytelling with graph knowledge and relational reasoning},
\newblock in: \bibinfo{booktitle}{Proceedings of the AAAI Conference on Artificial Intelligence}, volume~\bibinfo{volume}{35}, \bibinfo{year}{2021}, pp. \bibinfo{pages}{3022--3029}.
%Type = Article
\bibitem[{Chen(2025)}]{chen2025structured}
\bibinfo{author}{Y.-C. Chen},
\newblock \bibinfo{title}{Structured graph representations for visual narrative reasoning: A hierarchical framework for comics},
\newblock \bibinfo{journal}{arXiv preprint arXiv:2506.10008}  (\bibinfo{year}{2025}).
%Type = Inproceedings
\bibitem[{Mihalcea et~al.(2006)Mihalcea, Corley, Strapparava et~al.}]{mihalcea2006corpus}
\bibinfo{author}{R.~Mihalcea}, \bibinfo{author}{C.~Corley}, \bibinfo{author}{C.~Strapparava}, et~al.,
\newblock \bibinfo{title}{Corpus-based and knowledge-based measures of text semantic similarity},
\newblock in: \bibinfo{booktitle}{Aaai}, volume~\bibinfo{volume}{6}, \bibinfo{year}{2006}, pp. \bibinfo{pages}{775--780}.
%Type = Article
\bibitem[{Vuli{\'c} et~al.(2020)Vuli{\'c}, Ponti, Litschko, Glava{\v{s}}, and Korhonen}]{vulic2020probing}
\bibinfo{author}{I.~Vuli{\'c}}, \bibinfo{author}{E.~M. Ponti}, \bibinfo{author}{R.~Litschko}, \bibinfo{author}{G.~Glava{\v{s}}}, \bibinfo{author}{A.~Korhonen},
\newblock \bibinfo{title}{Probing pretrained language models for lexical semantics},
\newblock \bibinfo{journal}{arXiv preprint arXiv:2010.05731}  (\bibinfo{year}{2020}).
%Type = Inproceedings
\bibitem[{Amizadeh et~al.(2020)Amizadeh, Palangi, Polozov, Huang, and Koishida}]{amizadeh2020neuro}
\bibinfo{author}{S.~Amizadeh}, \bibinfo{author}{H.~Palangi}, \bibinfo{author}{A.~Polozov}, \bibinfo{author}{Y.~Huang}, \bibinfo{author}{K.~Koishida},
\newblock \bibinfo{title}{Neuro-symbolic visual reasoning: Disentangling},
\newblock in: \bibinfo{booktitle}{International Conference on Machine Learning}, \bibinfo{organization}{Pmlr}, \bibinfo{year}{2020}, pp. \bibinfo{pages}{279--290}.
%Type = Article
\bibitem[{Yi et~al.(2018)Yi, Wu, Gan, Torralba, Kohli, and Tenenbaum}]{yi2018neural}
\bibinfo{author}{K.~Yi}, \bibinfo{author}{J.~Wu}, \bibinfo{author}{C.~Gan}, \bibinfo{author}{A.~Torralba}, \bibinfo{author}{P.~Kohli}, \bibinfo{author}{J.~Tenenbaum},
\newblock \bibinfo{title}{Neural-symbolic vqa: Disentangling reasoning from vision and language understanding},
\newblock \bibinfo{journal}{Advances in neural information processing systems} \bibinfo{volume}{31} (\bibinfo{year}{2018}).
%Type = Inproceedings
\bibitem[{Li et~al.(2021)Li, Stengel-Eskin, Zhang, Xie, Tran, Van~Durme, and Yuille}]{li2021calibrating}
\bibinfo{author}{Z.~Li}, \bibinfo{author}{E.~Stengel-Eskin}, \bibinfo{author}{Y.~Zhang}, \bibinfo{author}{C.~Xie}, \bibinfo{author}{Q.~H. Tran}, \bibinfo{author}{B.~Van~Durme}, \bibinfo{author}{A.~Yuille},
\newblock \bibinfo{title}{Calibrating concepts and operations: Towards symbolic reasoning on real images},
\newblock in: \bibinfo{booktitle}{Proceedings of the IEEE/CVF International Conference on Computer Vision}, \bibinfo{year}{2021}, pp. \bibinfo{pages}{14910--14919}.
%Type = Article
\bibitem[{Khan et~al.(2025)Khan, Ilievski, Breslin, and Curry}]{khan2025survey}
\bibinfo{author}{M.~J. Khan}, \bibinfo{author}{F.~Ilievski}, \bibinfo{author}{J.~G. Breslin}, \bibinfo{author}{E.~Curry},
\newblock \bibinfo{title}{A survey of neurosymbolic visual reasoning with scene graphs and common sense knowledge},
\newblock \bibinfo{journal}{Neurosymbolic Artificial Intelligence} \bibinfo{volume}{1} (\bibinfo{year}{2025}) \bibinfo{pages}{NAI--240719}.
%Type = Inproceedings
\bibitem[{Mostafazadeh et~al.(2016)Mostafazadeh, Chambers, He, Parikh, Batra, Vanderwende, Kohli, and Allen}]{mostafazadeh2016corpus}
\bibinfo{author}{N.~Mostafazadeh}, \bibinfo{author}{N.~Chambers}, \bibinfo{author}{X.~He}, \bibinfo{author}{D.~Parikh}, \bibinfo{author}{D.~Batra}, \bibinfo{author}{L.~Vanderwende}, \bibinfo{author}{P.~Kohli}, \bibinfo{author}{J.~Allen},
\newblock \bibinfo{title}{A corpus and cloze evaluation for deeper understanding of commonsense stories},
\newblock in: \bibinfo{booktitle}{Proceedings of the 2016 Conference of the North American Chapter of the Association for Computational Linguistics: Human Language Technologies}, \bibinfo{year}{2016}, pp. \bibinfo{pages}{839--849}.
%Type = Article
\bibitem[{Ko{\v{c}}isk{\`y} et~al.(2018)Ko{\v{c}}isk{\`y}, Schwarz, Blunsom, Dyer, Hermann, Melis, and Grefenstette}]{kovcisky2018narrativeqa}
\bibinfo{author}{T.~Ko{\v{c}}isk{\`y}}, \bibinfo{author}{J.~Schwarz}, \bibinfo{author}{P.~Blunsom}, \bibinfo{author}{C.~Dyer}, \bibinfo{author}{K.~M. Hermann}, \bibinfo{author}{G.~Melis}, \bibinfo{author}{E.~Grefenstette},
\newblock \bibinfo{title}{The narrativeqa reading comprehension challenge},
\newblock \bibinfo{journal}{Transactions of the Association for Computational Linguistics} \bibinfo{volume}{6} (\bibinfo{year}{2018}) \bibinfo{pages}{317--328}.
%Type = Article
\bibitem[{Xu et~al.(2022)Xu, Wang, Yu, Ritchie, Yao, Wu, Zhang, Li, Bradford, Sun et~al.}]{xu2022fantastic}
\bibinfo{author}{Y.~Xu}, \bibinfo{author}{D.~Wang}, \bibinfo{author}{M.~Yu}, \bibinfo{author}{D.~Ritchie}, \bibinfo{author}{B.~Yao}, \bibinfo{author}{T.~Wu}, \bibinfo{author}{Z.~Zhang}, \bibinfo{author}{T.~J.-J. Li}, \bibinfo{author}{N.~Bradford}, \bibinfo{author}{B.~Sun}, et~al.,
\newblock \bibinfo{title}{Fantastic questions and where to find them: Fairytaleqa--an authentic dataset for narrative comprehension},
\newblock \bibinfo{journal}{arXiv preprint arXiv:2203.13947}  (\bibinfo{year}{2022}).
%Type = Inproceedings
\bibitem[{Choi et~al.(2021)Choi, On, Heo, Seo, Jang, Lee, and Zhang}]{choi2021dramaqa}
\bibinfo{author}{S.~Choi}, \bibinfo{author}{K.-W. On}, \bibinfo{author}{Y.-J. Heo}, \bibinfo{author}{A.~Seo}, \bibinfo{author}{Y.~Jang}, \bibinfo{author}{M.~Lee}, \bibinfo{author}{B.-T. Zhang},
\newblock \bibinfo{title}{Dramaqa: Character-centered video story understanding with hierarchical qa},
\newblock in: \bibinfo{booktitle}{Proceedings of the aaai conference on artificial intelligence}, volume~\bibinfo{volume}{35}, \bibinfo{year}{2021}, pp. \bibinfo{pages}{1166--1174}.
%Type = Inproceedings
\bibitem[{Tapaswi et~al.(2016)Tapaswi, Zhu, Stiefelhagen, Torralba, Urtasun, and Fidler}]{tapaswi2016movieqa}
\bibinfo{author}{M.~Tapaswi}, \bibinfo{author}{Y.~Zhu}, \bibinfo{author}{R.~Stiefelhagen}, \bibinfo{author}{A.~Torralba}, \bibinfo{author}{R.~Urtasun}, \bibinfo{author}{S.~Fidler},
\newblock \bibinfo{title}{Movieqa: Understanding stories in movies through question-answering},
\newblock in: \bibinfo{booktitle}{Proceedings of the IEEE conference on computer vision and pattern recognition}, \bibinfo{year}{2016}, pp. \bibinfo{pages}{4631--4640}.
%Type = Inproceedings
\bibitem[{Martin et~al.(2018)Martin, Ammanabrolu, Wang, Hancock, Singh, Harrison, and Riedl}]{martin2018event}
\bibinfo{author}{L.~Martin}, \bibinfo{author}{P.~Ammanabrolu}, \bibinfo{author}{X.~Wang}, \bibinfo{author}{W.~Hancock}, \bibinfo{author}{S.~Singh}, \bibinfo{author}{B.~Harrison}, \bibinfo{author}{M.~Riedl},
\newblock \bibinfo{title}{Event representations for automated story generation with deep neural nets},
\newblock in: \bibinfo{booktitle}{Proceedings of the AAAI Conference on Artificial Intelligence}, volume~\bibinfo{volume}{32}, \bibinfo{year}{2018}.
%Type = Inproceedings
\bibitem[{Fujimoto et~al.(2016)Fujimoto, Ogawa, Yamamoto, Matsui, Yamasaki, and Aizawa}]{fujimoto2016manga109}
\bibinfo{author}{A.~Fujimoto}, \bibinfo{author}{T.~Ogawa}, \bibinfo{author}{K.~Yamamoto}, \bibinfo{author}{Y.~Matsui}, \bibinfo{author}{T.~Yamasaki}, \bibinfo{author}{K.~Aizawa},
\newblock \bibinfo{title}{Manga109 dataset and creation of metadata},
\newblock in: \bibinfo{booktitle}{Proceedings of the 1st international workshop on comics analysis, processing and understanding}, \bibinfo{year}{2016}, pp. \bibinfo{pages}{1--5}.
%Type = Article
\bibitem[{Shvaiko and Euzenat(2011)}]{shvaiko2011ontology}
\bibinfo{author}{P.~Shvaiko}, \bibinfo{author}{J.~Euzenat},
\newblock \bibinfo{title}{Ontology matching: state of the art and future challenges},
\newblock \bibinfo{journal}{IEEE Transactions on knowledge and data engineering} \bibinfo{volume}{25} (\bibinfo{year}{2011}) \bibinfo{pages}{158--176}.
%Type = Article
\bibitem[{Lehmann et~al.(2015)Lehmann, Isele, Jakob, Jentzsch, Kontokostas, Mendes, Hellmann, Morsey, Van~Kleef, Auer et~al.}]{lehmann2015dbpedia}
\bibinfo{author}{J.~Lehmann}, \bibinfo{author}{R.~Isele}, \bibinfo{author}{M.~Jakob}, \bibinfo{author}{A.~Jentzsch}, \bibinfo{author}{D.~Kontokostas}, \bibinfo{author}{P.~N. Mendes}, \bibinfo{author}{S.~Hellmann}, \bibinfo{author}{M.~Morsey}, \bibinfo{author}{P.~Van~Kleef}, \bibinfo{author}{S.~Auer}, et~al.,
\newblock \bibinfo{title}{Dbpedia--a large-scale, multilingual knowledge base extracted from wikipedia},
\newblock \bibinfo{journal}{Semantic web} \bibinfo{volume}{6} (\bibinfo{year}{2015}) \bibinfo{pages}{167--195}.
%Type = Article
\bibitem[{Rajpurkar et~al.(2016)Rajpurkar, Zhang, Lopyrev, and Liang}]{rajpurkar2016squad}
\bibinfo{author}{P.~Rajpurkar}, \bibinfo{author}{J.~Zhang}, \bibinfo{author}{K.~Lopyrev}, \bibinfo{author}{P.~Liang},
\newblock \bibinfo{title}{Squad: 100,000+ questions for machine comprehension of text},
\newblock \bibinfo{journal}{arXiv preprint arXiv:1606.05250}  (\bibinfo{year}{2016}).
%Type = Article
\bibitem[{Chen and Jhala(2023)}]{chen2023customizable}
\bibinfo{author}{Y.-C. Chen}, \bibinfo{author}{A.~Jhala},
\newblock \bibinfo{title}{A customizable generator for comic-style visual narrative},
\newblock \bibinfo{journal}{arXiv preprint arXiv:2401.02863}  (\bibinfo{year}{2023}).
%Type = Article
\bibitem[{Chen and Jhala(2024)}]{chen2024collaborative}
\bibinfo{author}{Y.-C. Chen}, \bibinfo{author}{A.~Jhala},
\newblock \bibinfo{title}{Collaborative comic generation: Integrating visual narrative theories with ai models for enhanced creativity},
\newblock \bibinfo{journal}{arXiv preprint arXiv:2409.17263}  (\bibinfo{year}{2024}).

\end{thebibliography}
